
\input amstex
\magnification=1200
\baselineskip=16pt
\nopagenumbers
\TagsOnRight
\def\wh{\widehat}

\font\diez=cmr10

\def\ov{\overline}

\def\ep{\varepsilon}
\line{Preprint {\bf SB/F/94-223}}
\hrule
\vskip 0.5cm
\centerline{\bf TOPOLOGICAL SECTORS OF SPIN 1 THEORIES}
\centerline{\bf IN 2+1 DIMENSIONS}
\vskip 0.5cm
\centerline{\bf P\'{\i}o Jos\'e Arias ${}^{a,b}$}
\centerline{\bf and}
\centerline{\bf A. Restuccia ${}^{a}$}
\centerline{\it ${}^{a}$ Universidad Sim\'on Bol\'{\i}var, Departamento de
F\'{\i}sica}
\centerline{\it Apartado Postal 89000, Caracas 1080-A}
\centerline{\it Venezuela}
\centerline{\it ${}^{b}$Departamento de F\'{\i}sica, Facultad de Ciencias}
\centerline{\it Universidad Central de Venezuela, AP 20513, Caracas 1020-a}
\centerline{\it Venezuela}
\centerline{\it e-mail:  parias{\@}usb.ve, arestu{\@}usb.ve}
\vskip 1cm
\centerline{\it Dedicated to the memory of Professor Carlos Aragone }
\centerline{\it founder of the Group of Relativity and Fields at USB}
\vskip 0.5cm

\noindent
{\bf ABSTRACT}
{\diez
{\narrower\flushpar
It is shown that the Topological Massive and ``Self-dual'' theories,
which are known to provide locally equivalent descriptions of spin 1 theories
in 2+1 dimensions, have different global properties when formulated over
topologically non-trivial regions of space-time. The partition function
of these theories, when constructed on an arbitrary Riemannian
manifold, differ by a topological factor, which is equal to the partition
function of the pure Chern-Simons theory. This factor is related to
the space of solutions of the field equations of the Topological
Massive Theory for which the connection is asymptotically flat but
not gauge equivalent to zero. A new covariant, first order,
gauge action,which generalize the ``Self-dual'' action, is then proposed. It
is obtained by sewing local self-dual theories. Its global equivalence to the
Topological Massive gauge theory is shown.\par}}

\vskip 0.5cm

\hrule
\vskip 0.5cm
\centerline{\bf UNIVERSIDAD SIMON BOLIVAR}

\newpage

Vector and tensor gauge theories [1-5] in three dimensional space time enjoy
very special properties arising from gauge invariant, topologically non
trivial terms which provide masses for the gauge fields. This topological
terms are related to the Chern-Simons characteristic classes, which may be
obtained from four dimensional Pontryagin invariants and also arise
naturally  from the four dimensional topological $BF$ theory [6,7]. These
three topological functionals: Chern-Simons, Pontryagin and $BF$ actions are
just  the starting point for most of the Topological Quantum Field Theories
[6,8-11].We are going to discuss one of this properties, enjoyed
by the  three dimensional vector gauge field theories, related to the
topological  terms mentioned above.

It is known that the spin 1 theory in $2+1$ dimensions may be
described by two covariant actions: The Topological Massive $(TM)$
gauge action[2] and the, first order, Self Dual $(SD)$ action[12,13]. Also,
it has been shown that the $SD$ action corresponds to a gauge fixed
version of the $(TM)$ theory [14-15]. This gives rise to the possibility
of constructing two different covariant effective actions from the
same gauge theory.  One of this covariant gauge
fixed actions is of second order in derivatives and arises  by imposing the
Lorentz covariant gauge fixing condition which involves, as  usual, a
Lagrange multiplier of the canonical formulation. The other one, the
$SD$ formulation, results from a canonical gauge fixing procedure
which is not Lorentz covariant, however the full action can be
rewritten in a Lorentz covariant form[15]. According to the $BFV$ arguments,
the partition functions and, in general,
the  physical observables are independent of the gauge fixing conditions
within  the admissible set. Under assumption of simply connectness of the
base manifold everything is equivalent [14-16], however once we consider
both effective actions over a  topologically non trivial base manifold, where
the topological Chern-Simons  structure may contribute to the observables of
the theory, the relation  between both formulations is not of a trivial
equivalence. In this case, only the gauge fixed action with the Lorentz term
preserves the global properties of the original gauge theory. However, both
theories describe the same propagating physical degrees of freedom and have
the same local properties. This is a novel feature not enjoyed by any known
field theorie in four dimensions.

We are going to  explicity show this global difference by
comparison of the partition functions associated to both actions when
formulated over a Riemannian base manifold and show that they differ by a
 topological factor equal to the partition function of the
 pure Chern-Simons action, which, as is well known, may be expressed in terms
of the topological Ray-Singer torsion. This  topological factor has its
origin in the difference between the space of solutions  of the field
equations associated to both theories. In fact, the space of flat
connections is a solution of the field equations for the $TM$  gauge theory,
while the only flat connection which is a solution of the $SD$ theory is the
trivial one. For simply connected regions of space-time both spaces of
solutions are  identical but for topologically non trivial base manifolds
the space of flat  connections admits non-trivial configurations. Gauge
inequivalent flat connections are characterized by their holonomy around
non-contractible loops. This is equivalent to specifying homomorphisms
from the fundamental $\Pi_1$ group into the structure group ($U(1)$ in
our case). The topological index
 $$ I=\oint a , \tag 1 $$
evaluated for asimptotically flat solutions, is zero in the case of the
$SD$ theory and is $\neq 0$ for the non trivial solutions of the $TM$
gauge theory (explicit solutions have been found in [4,5,16]). This
space of asymptotically flat  solutions correspond exactly to the
Chern-Simons ($CS$) classical solutions, which are connected with the
description of anyons [17,18] ( see the reviews in [17] and the references
there in ), and because of this reason the solutions with non trivial
topological index $I$ are said to have ``anyonic behaviour''.

After showing this global difference in the two formulations, we are going
to present a new covariant, first order, gauge action which generalizes
the $SD$ one and is locally and globally equivalent to the $TM$ action.

Let us start our discussion by showing briefly the canonical equivalence
between both theories, over a simply connected region of space-time. The
Lagrangian density of the $SD$ theory is
$$
L=\frac{m^2}{2}a_\mu a^\mu -\frac{m}{2}\ep^{\mu\nu\rho}a_\mu \partial_\nu
a_\rho ,\tag 2
$$
and the canonical Hamiltonian density associated to it is given by
$$
\Cal H _0=\frac{m^2}{2}a_ia_i +\frac{1}{2}(\ep_{ij}\partial_i a_j)^2 \tag 3
$$
subject to two second class constraints
$$
\theta_i =P_i - \frac{m}{2}\ep_{ik}a_k =0 \tag 4
$$
where $P^i$ is the conjugate momenta associated to $a_i$ (we are using
metric signature $(+--)$). It was noticed, in  [15], that they may be
interpreted as a first class constraint $\theta$ and its  associated gauge
fixing condition $\chi$, $$\align
\theta &=-\partial_i\theta_i=-\partial_iP_i+\frac{m}{2}\ep_{ij}\partial_ia_j
\tag 5 \\
\chi &=\ep^{ij}\partial_i\theta_j=\ep_{ij}\partial_iP_j+\frac{m}{2}
\partial_ka_k. \tag 6
\endalign
$$
The system (2) may, then, be considered as a gauge theory governed by the
Hamiltonian density
$$
\wh{\Cal H}_0=\frac{1}{2}P_kP_k+\frac{m}{2}\ep_{ij}P_ia_j+
\frac{m^2}{8}a_ka_k+ \frac{1}{2}(\ep_{ij}\partial_ia_j)^2, \tag 7
$$
(which reduces to (3), under (5) and (6)) subject to the first class
constraint  $\theta$.

The quantized formulation of this new gauge system is equivalent to
that of the original one [15]. The  new Hamiltonian density (7) and the first
class constraint $\theta$, are just the  ones which emerge from the canonical
analysis of the $TM$  gauge theory. Moreover, with $\chi$ as a gauge fixing
condition the effective action is just the $SD$ action.
It is important to notice that (5) and (6) are equivalent to (4) provided
the region of space time we are considering is simply connected. So both
theories, the $SD$ and the $TM$, are completely equivalent at the
classical and quantum level on a simply  connected region of space-time.

We now compare the partition function of the two theories. The
global difference in the space of classical solutions is going to be
reflected in the evaluation of the partition functions where a topological
factor will arise, in the $TM$ case, which is not present in the $SD$
 one. In order to detect this topological factor we consider the
formulation of both theories in a general Riemannian 3-manifold
background, $M$. We start considering the canonical formulation with the
correct  quantum measure and evaluate the partition function after
integration of the  conjugate momenta. At the end of the paper we will
discuss briefly the zero modes contributions.

We have that for the $SD$ theory
$$
S^{SD}=\int_M d^3x\sqrt{g}a_\mu T^{\mu\nu}a_\nu \tag 8
$$
with
$$
T^{\mu\nu}=\frac{m}{2}(mg^{\mu\nu}-\frac{1}{\sqrt{g}}\ep^{\mu\lambda\nu}
\Cal{D}_\lambda)\tag 9
$$
where $\Cal{D}_\lambda$ is the covariant derivative, on the manifold. The
partition function is then
$$
Z_{SD}=\rho (\det \sqrt{g}T)^{-1/2} \tag 10
$$
where $\rho =(\det (m^2\delta ))^{1/2}$ arises from the Senjanovic-Fradkin
measure term $\det^{1/2}\{\theta_i,\theta_j\}$.

For the $TM$ theory, in the Lorentz gauge, the $BRST$ invariant effective
action, on $M$, takes the  form
$$
S^{TM}_{eff}=\int_Md^3x\sqrt{g}[a_\mu S^{\mu\nu}a_\nu -mB\Cal{D}_\mu
a^{\mu}-m\ov{C}\Cal{D}_\mu \partial^\mu C]\tag 11
$$
where
$$
S^{\mu\nu}=\frac{1}{2}[(g^{\mu\nu}\Cal{D}_\lambda \Cal{D}^\lambda -\Cal{D}^\nu
\Cal{D}^\mu)+\frac{m}{\sqrt{g}}\ep^{\mu\lambda\nu}\Cal{D}_\lambda ]\tag 12
$$
and $B,C,\ov{C}$ are, respectively the Lagrange multiplier, ghost and
antighost fields introduced when the canonical $BRST$ procedure is applied.
The  degenerancy of $S^{\mu\nu}$ is easily observed because
$S^{\mu\nu}\partial_\nu \lambda =0$ for any 0-form $\lambda$. If we define
the differential operator $$
C^{\mu\nu}=\frac{1}{m}\ep^{\mu\nu\lambda}\partial_\nu \tag 13
$$
which is proportional to the kinetic operator of the pure Chern-Simons theory,
is inmediate to see that acting on 1-forms
$$
T^{\mu\nu}\frac{1}{\sqrt{g}}C_{\nu}{}^\lambda =\frac{1}{\sqrt{g}}C^{\mu\nu}
T_{\nu}{}^\lambda =S^{\mu\lambda} \tag 14
$$
this fact is going to be crucial when we compare the two partition
functions.  Before considering that point, we must notice that in this
Riemannian  3-manifold $M$, one can define the Hodge dual * which maps
p-forms on (3-p)-forms  and satisfies **={\bf 1}. The adjoint of the exterior
derivate $d$ is  $\delta =(-1)^{p}*d*$ when acting on p-forms, and satisfies
$\delta^2=0$.  Finally the Laplacian on p-forms (the Laplace-Beltrami
operator) is defined to be  $\Delta\equiv \delta d+d\delta$, as usual. So
over a 0-form $\lambda$ $$
\Delta_0 \lambda =-\Cal{D}_\mu\partial^\mu \lambda \tag 15
$$
and over a l-form $V_\mu$
$$
\Delta_1 V_\mu =-\Cal{D}_\nu \Cal{D}^\nu V_\mu +R_{\mu\nu}V^\nu \tag 16
$$
where $R_{\mu\nu}$ is the Ricci tensor.

The action can , then, be rewritten as
$$
S^{TM}_{eff}=\int_M d^3x\sqrt{g}[\Phi^tK_B\Phi +m\ov{C}\Delta_0 C]\tag 17
$$
where
$$
K_B\equiv \left(\matrix
S^{\mu\lambda} & \frac{m}{2}\partial^\mu \\
-\frac{m}{2}\Cal{D}^\lambda & 0
\endmatrix \right)
\tag 18
$$
and $\Phi^t = (a_\lambda {\quad}B)$. Then, the partition function will
be[11]  $$
Z_{TM}=\det (\sqrt{g}K_B)^{-1/2}\det m\delta \det \Delta_0. \tag 19
$$

To evaluate the determinat of $K_B$ we take the square the operator which
is diagonal
$$
K_B^2=\left(\matrix
S^{\mu\alpha}S_\alpha{}^\lambda -\frac{m^2}{4}\partial^\mu\Cal{D}^\lambda & 0
\\
0 & -\frac{m^2}{4}\Cal{D}_\alpha \partial^\alpha
\endmatrix \right) \tag 20
$$
For the part which act on 1-forms, it can be seen, using (14), that
$$
S^{\mu\alpha}S_\alpha{}^\lambda -\frac{m^2}{4}\partial^\mu\Cal{D}^\lambda =
\frac{1}{m^2}T^{\mu\theta}(-g_{\theta\rho}\Cal{D}_\alpha \Cal{D}^\alpha +
\Cal{D}_\rho \Cal{D}_\theta -\Cal{D}_\theta \Cal{D}_\rho ) T^{\rho\lambda},
\tag 21
$$
so
$$
K_B^2=\left(\matrix
\frac{1}{m^2}T\Delta_1T & \\
 & \frac{m^2}{4}\Delta_0
\endmatrix \right)
$$
and
$$
Z_{TM}=\rho (\det \sqrt{g}T)^{-1/2}(\det \Delta_1)^{-1/4}(\det \Delta_0)^{3/4}.
\tag
22
$$
Here $\rho$ is the same factor as in the $SD$ theory. The two partition
functions differ, then, in a factor which is just the partition function of
the pure Chern-Simons theory. This factor is related to the Ray-Singer
torsion, $T(M_3)$, by  $Z_{CS}=T(M_3)^{-1/2}$, and is metric independent
[19,8]. For an even dimensional oriented compact manifold, without
boundary,  $T(M_{2m})=1$. This fact is obtained, from the scaling invariance
in path  integrals of some $BF$ systems [8]. In odd dimensions,
in distinction, those invariances, together with the Hodge duality property
$(*\Delta = \Delta *)$, lead to identities that give no information about
$T(M_{2m+1})$ [8]. In other direction, the two point function of
Topological  Field theories, whose partition function is the Ray-Singer
torsion, can be  used as a definition of generalized linking number between
surfaces [7],  which is connected with the concepts of fractional statistics
[17,18,20].

The  common factor between $Z_{SD}$ and $Z_{TM}$ could be expected because
the  self-dual equations of motion constitutes a minimal realization
of the  ``Pauli-Lubanski'', and mass shell conditions for the spin 1
representations  of the Poincar\'e group in 2+1 dimensions [21]. The extra
factor is connected  with the ``topological'' properties of the $TM$ theory,
and explains why, for the $SD$ theory, there is no ``anyonic behaviour''.
This topological factor reduces to one, for a simply connected region of
space time, where both theories are equivalent. In fact, assuming
$M_3=R$x$M_2$ and $M_2$ simply connected we have
$$
\det \Delta_1(M_3)=\det \Delta_1(M_2)\det\Delta_0 . \tag 23
$$
Now, using Proposition 4 in [8], it can be seen that
$\det\Delta_1(M_2)=(\det\Delta_0)^2$; hence
$$
\det\Delta_1(M_3)=(\det\Delta_0 )^3, \tag 24
$$
and
$$
(\det \Delta_1)^{-1/4}(\det \Delta_0)^{3/4}=1. \tag 25
$$
(10) and (22), show that the $TM$ theory, which is locally equivalent to the
$SD$ theory, is globally different to it.

When the base manifold is $R$x$M_2$, the occurence of $M_2$ as a multiply
connected manifold arises in various interesting models. The simplest one is
when we couple minimally the $TM$ or the $SD$ theories to a source that
consists of a charge particle at the origin. If the source has ``dipole
strengh'' $\sigma$, the static solutions, outside sources, for the
$SD$, $TM$ and pure $CS$ theories, are related by [4,16]
$$
a_0^{CS}=0,\tag 26,a
$$
$$
a_0^{SD}=a_0^{TM}=-(q+m\sigma)Y(mr),\tag 26,b
$$
$$
\align
a_i^{TM}&=a_i^{SD} + a_i^{CS}\\
&=\frac{q+m\sigma}{m}\epsilon_{ij}\partial_jY(mr)+(-\frac{q}{m}\epsilon_{ij}
\partial_jC(mr)+\partial_j\lambda),\tag 26,c
\endalign
$$
where the longitudinal part of $a_i$ remains unfixed, in the $TM$ and pure
$CS$ theories, because of the gauge freedom. $Y(mr)$ and the $C(mr)$ are,
respectively, the Yukawa and Coulomb Green functions, i.e. $(-\Delta
+m^2)Y(mr)=(-\Delta)C(mr)=\delta^2 (r)$. Asymptotically $a_\mu^{TM}\sim
a_\mu^{CS}$ and $a_\mu^{SD}\sim 0$. In the special case that
$q+m\sigma=0$, these last relations hold everywhere [4]. For both cases
$F_{\mu\nu}=0$ but only for the $TM$ and $CS$ theories the
potential $a_\mu$ is closed and not exact. More precisely, the index $I$,
(1), (for loops around the origin) becomes $q/m$ as in the pure $CS$
theory. This result is used to implement fractional statistics
dynamically [14, 16-18].

The local relation with the $SD$ solutions arises because $a_i^{TM}$ can be
rewritten as $(q+m\sigma=0)$
$$
a_i^{TM}=a_i^{CS}=\partial_i(\lambda-\frac{q\Theta}{2\pi m}),\tag 27
$$
where $\Theta=arctg(x_2/x_1)$ is a multivalued function. However,
$\partial_i\Theta$ is a well defined 1-form on the punctured plane known as
an Oersted-Amper 1-form [22]. This 1-form is closed $(F_{ij}=0)$, but not
exact ($\oint\partial_{i}\Theta dx^i\neq 0$, for loops around the
origin). The possibility of fixing gauge in such a way that $a_i=0$ can only
be performed on simply connected regions, but not globally (this gauge is
commonly known as the singular gauge that eliminates the potential). So the
non equivalence between both theories is also reflected in this special
example. It can be shown that, by making a different kind of coupling, the
self dual solutions can reproduce the $TM$ ones, but it must be a non-local
type of coupling [14,16]. If we want to obtain the topological sector of the
space of solutions not present in the $SD$ model it seems that one should
consider patching and sewing ``$SD$ formulations'' over simply connected
sectors of the base manifold . In order to do so, we start considering the
functional integral of the $TM$ theory. Its functional measure is
 $$
\delta^2(\theta)\delta ^2(\chi)det\{\theta,\chi\}, \tag 28
$$
where $\theta$ and $\chi$ are given by (5) and (6) respectively. This may be
rewritten as
$$
\langle\quad\prod_{i=1}^2 \delta^2(\theta_i
+m\epsilon_{ij}\omega_j)det^{1/2}\{\theta_i
,\theta_j\}\mu\quad\rangle_{H_1},\tag 29
$$
where $\omega_i dx^i$ is a 2 dimensional closed 1-form satisfying
$\partial_i\omega_i =0$. This condition only fixes the exact forms
corresponding to a given cohomology class. Integration is done on the space
of cohomology classes, $H_1$ with measure $\mu = Z_{CS}(R$x$M_2)$. $\theta_i$
is defined as in (4). (29) constitutes the generalization of the arguments,
used from (2) through (7), to a topological non trivial space.

After integration on the conjugated momenta we arrive to the functional
integral associated to the following action
$$
S'=\int d^3 x[\frac{m^2}{2}(a_\mu +\omega_\mu)(a^\mu
+\omega^\mu)-\frac{m}{2}(a_\mu
+\omega_\mu)\epsilon^{\mu\nu\lambda}\partial_\nu(a_\lambda
+\omega_\lambda)],\tag 30
$$
where the closed form $\omega_\mu dx^\mu$ satisfies the gauge condition
$\partial^\mu\omega_\mu =0$. $a_\mu$ and $\omega_\mu$ are
independent fields. Functional integration on $\omega_\mu$ is performed
in $H_1$ with measure $\mu$. The functional integral of the $TM$ theory is
then equivalent to the functional integral associated to (30) which is also a
gauge invariant action. The condition imposed to $\omega_\mu$ can always be
selected on any cohomology class.In the previous
argument we assumed $M_3=M_2$x$R$, in order to perform the canonical
analysis. In the particular case that $M_3$ is simply connected
$\omega_\mu =0$ the usual $SD$ formulation is regained.

The constraints associated to (30) are
$$
P_i -\frac{m}{2}\epsilon_{ij}(a_j +\omega_j)=0,\tag31
$$
and
$$
\langle\pi^I -P^j\Lambda_j^I\rangle_{M_2}=0,\tag32
$$
here $P^j$ has the same meaning as before, while $\pi^I (t)$ is the
conjugate momenta associated to $\alpha^I (t)$, where $\Lambda_i^I(x)$ is a
basis of closed 1-forms, so
$$
\omega_i (t,x)=\alpha^I (t)\Lambda_i^I(x). \tag33
$$
Lastly $\langle\qquad\rangle{}_{M_2}$ denotes integration on $M_2$. We may
fix the gauge transformations generated by (32) taking
$$
\alpha^{I}(t) = constant. \tag 34
$$
then the condition on  $\omega_\mu$ reduces to $\partial_i\omega_i =0$,
which was the restriction obtained in (29).

The classical field equations arising from (30) are
$$
mA^\mu -\epsilon^{\mu\nu\lambda}\partial_\nu a_\lambda = -m\omega^\mu,\tag
35 $$
which may be rewritten as
$$
\epsilon_{\rho\gamma\mu}\partial^\gamma [
ma^\mu -\epsilon^{\mu\nu\lambda}\partial_\nu a_\lambda]=0.\tag 36
$$
(37) being the classical equations of the $TM$ theory. Variations with
respect to the $\alpha^{I}$ do not give any new equation of motion. The
partition functions associated to the $TM$ and the modified $SD$
theories ( in (30)) are also equal as we have mentioned. This feature may be
shown from the analysis of the functional measure as we did, or by direct
evaluation as in (10) and (22) where the volume of the zero modes of the $CS$
operator must now be included. Details of this analysis will be reported
elsewhere.

We have shown that the $TM$ and $SD$ theories, which are locally
equivalent, have different global properties. The difference arises, as
is known, at the classical level where non-trivial flat connections are
solutions of the $TM$ theory, while the trivial flat connection is the only
admissible solution for the $SD$ theory. We observe , then, by explicit
evaluation, that the partition functions differ by a topological factor, the
$CS$ partition function associated to that topological sector of the space
of solutions. Finally, we have constructed a covariant extension of the $SD$
theory, also of first order in derivatives, which is exactly
equivalent, locally and globally, to the $TM$ theory.

We have consider only the spin  1 abelian
theory. We expect analogous results for other spins in $3-D$. The case of
spin 2 linear theories is particulary interesting since there are
three equivalent linear theories with the same local physics [2,23] but
clearly with different global properties [4,5]. Also there is a kind of
factorization analogous to (14) and a gauge fixing procedure [24],
connecting one theory to the other. This will be reported elsewhere.

\vskip 1cm

\item{}{\bf References}

\item{1-}J. Schonfeld, Nuc. Phys. {\bf B185} (1981) 157.
\item{2-}S. Deser, R. Jackiw and Tempelton, Phys. Rev. Lett. {\bf 48} (1982)
975; Ann. Phys. {\bf 140} (1982) 372; (E) {\bf 185} (1988) 406.
\item{3-}R. Jackiw, Nuc. Phys. {\bf B252} (1985) 343.
\item{4-}S. Deser, Phys. Rev. Lett. {\bf 64} (1990) 611.
\item{5-}C. Aragone and P. J. Arias, {\it ``More gravitational anyons''}, in
the Proceedings of SILARG VIII (1993) World Scientific, in press.
\item{6-}G. T. Horowitz, Commun. Math. Phys. {\bf 125} (1989) 417.
\item{7-}G. T. Horowitz and M. Srednicki, Commun. Math. Phys. {\bf 130}
(1990) 83.
\item{8-}M. Blau and G. Thompson, Ann. Phys. {\bf 205} (1991) 130; Phys.
Lett. {\bf B255} (1991) 535.
\item{9-}A. S. Schwarz, Commun. Math. Phys. {\bf 67} (1979) 1.
\item{10-}M. I. Caicedo and A. Restuccia, Phys. Lett. {\bf B307} (1993) 77.
\item{11-}E.Witten, Commun. Math. Phys. {\bf 117} (1988) 353.
\item{12-}P. K. Townsend, K. Pilch y P. van Nieuwenhuizen, Phys. Lett.
{\bf B136} (1984) 38.
\item{13-}C. R. Hagen, Ann. Phys. {\bf 157} (1984) 342.
\item{14-}S. Deser and R. Jackiw, Phys. Lett. {\bf B139} (1984) 371.
\item{15-}R. Gianvittorio, A. Restuccia and J. Stephany, Mod. Phys. Lett.
{\bf A6} (1991) 2121.
\item{16-}P. J. Arias, Doctoral Thesis, U.S.B. (1994).
\item{17-}R.Mackenzie and F.Wilczek, Int. J. Mod. Phys. {\bf A3} (1988)
2827; S.Forte, Rev. Mod. Phys. {\bf 64} (1992) 517.
\item{18-}G.W.Semenoff, Phys. Rev. Lett. {\bf 48} (1988) 517.
\item{19-}D. B. Ray and I. M. Singer, Adv. Math {\bf 7} (1971) 145.
\item{20-}F. Wilczek and A. Zee, Phys. Rev. Lett. {\bf 51} (1983) 2250; A.
M.  Polyakov, Mod. Phys. Lett. {\bf A3} (1988) 325.
\item{21-}R. Jackiw and V. P. Nair, Phys. Rev. {\bf D43} (1991) 1933.
\item{22-}C.Von Westenholz, {\it ``Differential Forms in Mathematical
Physics''},North Holland Publishing Company (1981) pp. 295.
\item{23-}C.Aragone and A. Khoudeir,Phys.Lett.{\bf B173} (1986) 141.
\item{24-}P.J.Arias and J. Stephany,{\it ``Gauge invariance and second
class constraints in 3D linearized massive gravity''}, Preprint SB/F/93-210
(U.S.B.).
 \bye